\documentclass[11pt,twoside]{article}
\usepackage{asp2006}
\usepackage{psfig}
\usepackage{epsf}
\usepackage{graphics}
\usepackage{lscape}
\def\edcomment#1{\iffalse\marginpar{\raggedright\sl#1\/}\else\relax\fi}
\reversemarginpar

\begin{document}
\title{The Outer Cut--Off of the Giant Planet Population and the 6pc--Survey}
\author{D. Apai, M. R. Meyer, P. Hinz}
\affil{Steward Observatory, The University of Arizona, 933 N. Cherry Avenue, AZ--85721}
\author{M. Kasper}
\affil{European Southern Observatory, Karl--Schwarzschild--Str. 2, 85748 Garching, Germany }

\begin{abstract}
We present results from two high--contrast imaging surveys that exploit a novel technique, L--band angular differential 
imaging. Our first survey targeted 21 young stars in the $\beta$~Pic and Tuc--Hor moving groups with VLT/NACO reaching 
typical sensitivities  of $<$1~M$_{\mathrm Jup}$ at $r>20$~AU. 
The statistical analysis of the null result demonstrates that the giant planet population is truncated at 30~AU or less (90\% confidence level).
Our second, on--going MMT/Clio survey utilizes the unique sensitivity achieved in the L--band for old planets
to probe all M--dwarf stars within 6~pc. The proximity of these targets enables direct imaging of planets in orbits like Jupiter for the first time --- a
key step for directly imaging giant planets.
\end{abstract}

\vspace{-0.5cm}
\section{General Introduction}

The study of exoplanetary systems is arguably the most rapidly developing field in modern astrophysics. Surprisingly, much
progress has been made without directly imaging a single planet: radial velocity/microlensing and primary and secondary planet eclipses provide limited, but valuable insights. Direct imaging of planetary systems will have a fundamental impact on the field --- a single, low--resolution 3--5 $\mu$m spectrum of a planet may carry more information than all existing Spitzer transit  photometry combined. As was the case in the search for the first brown dwarf, or for radial velocity and planet transit techniques, achieving the first firm detection is a very difficult and often frustrating challenge. But these investments paid off rapidly by opening whole new classes of objects for study.

\section{L--band Angular Differential Imaging}

Using rotational subtraction of the persistent speckle pattern and the image artifacts, adaptive optics systems
are capable of reaching very high contrast. Low thermal--background telescopes, such as  Very Large Telescope (VLT) and the 
Multiple Mirror Telescope (MMT),  can utilize this technique in the L--band where the superior adaptive optics--correction 
(Strehl ratios 80\%--90\% vs. 30--45\% in H--band)  further increases the contrast. Our recent VLT/NACO 
L--band observations routinely achieved $\Delta$L=11 mag contrast at 1" and larger  separations. This contrast, in combination 
with the typical  color of  older planets (H$-$L=5.5 mag, 2 Gyr, 5 M$_{Jup}$, \citealt{Baraffe2003}), provides  more than 1~mag 
sensitivity increase over spectral differential imaging  \citep{Kasper2007}.

\section{The Scarcity of Giant Planets at Large Separations}

Using the L--band angular differential imaging technique on the VLT/NACO we carried out a survey of 
21 young stars, members of the nearby Tucana--Horologium (10--30 Myr) and the $\beta$ Pictoris moving groups ($\sim$12 Myr). 
No substellar companions were found around the target stars, but the companion to 51 Eri, GJ 3305, was 
found to be a very close binary on an eccentric orbit. Our sensitivity would have allowed the detection of 
companions as small as a Jupiter mass at orbital distances typically of 5 AU. 

The absence of detected companions sets constraints on the frequency and maximum orbital distance of 
giant exoplanets. We show that a radial distribution of planetary companion with a maximum 
orbital radius exceeding 30 AU  --- in combination with a power--law index of 0.2 --- can be rejected at a 
90\% confidence level (see, Fig. 1 and \citealt{Kasper2007}). This demonstrates that giant planets are
relatively rare at large separations. A similar conclusion was reached subsequently by other independent  surveys
(e.g. \citealt{Lafreniere2007}).

\section{Imaging Planets on the Scale of our Inner Solar System}

The result of our NACO survey also offers an explanation of why most current and past 
direct imaging surveys failed to detect a giant planet population: 
most of these focus on near--infrared wavelengths (1--2.2 $\mu$m), 
representing a trade--off between the rise of the thermal background toward the longer wavelengths and the adaptive optics 
performance degrading toward the shorter wavelengths. 
{\em However, the fact that the 1--2.2 $\mu$m flux of giant planets rapidly declines with time limits the age of the ideal target 
stars to less then 30--50 Myr.} Because young stars are typically at 40 pc or beyond, even the 
highest-order adaptive optics-based systems can only probe the outskirts ($>$1" or $>$40 AU) of these exoplanetary systems, 
where giant planets are rare \citep{Kasper2007}.

The key novelty of the L--band angular differential imaging is that it is also sensitive to planets as old as a 
few Gyr. This enables the study of relatively old nearby stars, whose proximity allows us to directly planets on orbits comparable
to that of  Jupiter ($r>3$~AU).

Exploiting the strength of this technique we are carrying out a 6~pc volume--limited survey of M--stars 
using the MMT/Clio 3--5 camera (e.g. \citealt{Hinz2006}). With 14 allocated nights we are covering
35 northern M--dwarfs within 6~pc. The M--dwarf survey will complement the Sun--like star survey of
Heinze et al. (in prep.)  surveying the complete stellar population within 6 pc. 
The survey is currently about 2/3 complete and the data reduction and analysis is ongoing  (Apai et al., in prep.).

 \setcounter{figure}{0}
 \begin{figure}[!ht]
 \plotfiddle{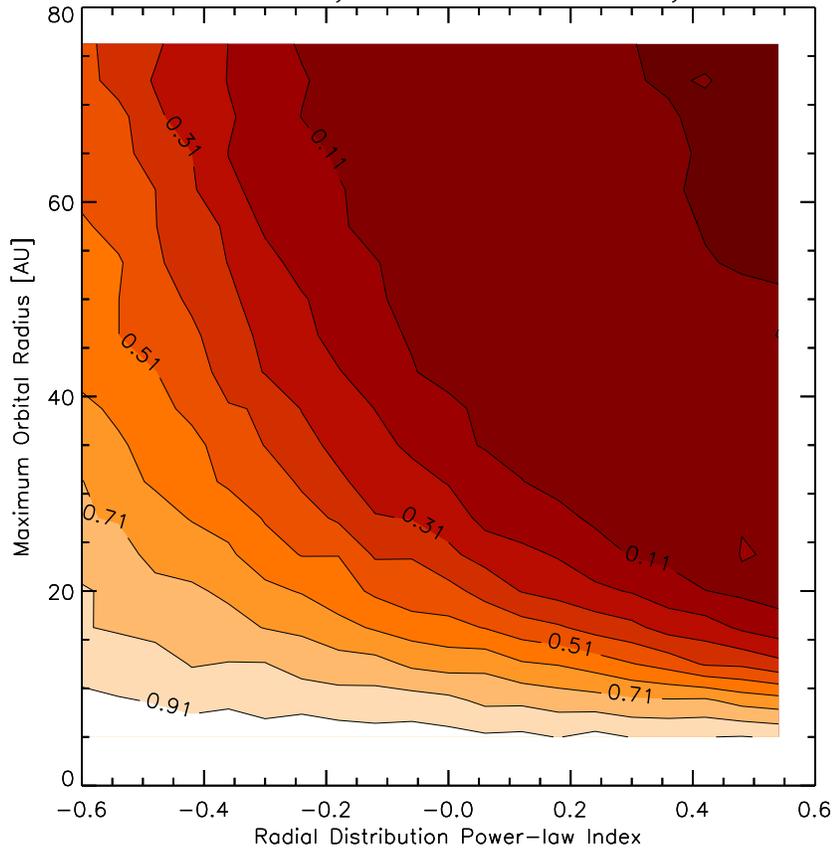}{11cm}{0}{60}{60}{-180}{0}
 \caption{Map of probability that the planet population simulated for a given $\mathrm dn/da \sim a^{-\alpha} $ and $r_{max}$ value is consistent with only 
 non---detections in our VLT/NACO survey.}
 \end{figure}

\section{Conclusions}
We present results from a novel high--contrast imaging technique. Our NACO survey of 21 nearby young stars 
demonstrates that the giant planet population does not extend beyond 30 AU and suggests a cut-off at radius $<$ 15 AU.
Most previous imaging surveys have not detected planets because they targeted young stars ($>$ 20 pc) 
forcing them to probe orbital radii $>$20~AU. 
Our ongoing MMT 6pc volume--limited survey is probing --- for the first time --- the massive giant planet population 
around the closest stars with orbital radii $>$3.

\end{document}